\documentclass[aps, prl, twocolumn, showpacs, floatfix,10pt,superscriptaddress]{revtex4-2}
\usepackage{amsmath}
\usepackage{amsfonts}
\usepackage{amsbsy}
\usepackage{amssymb}
\usepackage{graphicx}
\usepackage{textcomp}
\usepackage[caption=false,singlelinecheck=false]{subfig}
\usepackage{xcolor}
\usepackage{mathrsfs}
\usepackage{mathtools}
\usepackage{bm}
\usepackage{gensymb}
\usepackage{braket,wasysym}
\usepackage{float}
\usepackage{tikz}
\usepackage[colorlinks,linkcolor=blue,citecolor=red,filecolor=magenta,urlcolor=red,breaklinks]{hyperref}
\usepackage[bottom]{footmisc}
\usepackage{verbatim}

\hypersetup{colorlinks=true, urlcolor=blue, citecolor=red, pdfborder={0 0 0}}
\usepackage{breakurl}
\usepackage{natbib}

\allowdisplaybreaks

\newcommand{\tcb}[1]{\textcolor{blue}{#1}}

\normalfont
\def\be{\begin{equation}}
	\def\ee{\end{equation}}
\def\bea{\begin{eqnarray}}
	\def\eea{\end{eqnarray}}

\begin{document}
\title{
 Quantum Hyperuniformity and Quantum Weight
}

\author{Junmo Jeon}
\email{junmojeon@sophia.ac.jp}
\author{Shiro Sakai}
\email{shirosakai@sophia.ac.jp}
\affiliation{Physics Division, Sophia University, Chiyoda-ku, Tokyo 102-8554, Japan}

\date{\today}
\begin{abstract}
Extending hyperuniformity from classical to quantum fluctuations in electron systems yields a framework that identifies quantum phase transitions and reveals underlying gap structures through the quantum weight. We study long-wavelength fluctuations of many-body ground states through the charge-density structure factor by incorporating intrinsic quantum fluctuations into hyperuniformity. Although charge fluctuations at zero temperature are generally suppressed by particle-number conservation, their long-wavelength scaling reveals distinct universal behaviors that define quantum hyperuniformity classes. By exemplifying the Aubry–André model, we find that gapped, gapless, and localized–critical–extended phases are sharply distinguished by the quantum hyperuniformity classes.
Notably, at the critical point, multifractal wave functions generate anomalous scaling 
behavior. 
We further show that, in quantum-hyperuniform gapped phases, the quantum weight provides a quantitative measure of the gap size through a universal power-law scaling. Along with classical hyperuniformity, quantum hyperuniformity serves
a direct fingerprint of quantum criticality and a practical probe of quantum phase transitions in aperiodic electron systems.
\end{abstract}
\maketitle

\textit{\tcb{Introduction---}}
Quantum phase transitions are traditionally characterized by symmetry breaking and local order parameters~\cite{cohen2016fundamentals,sondhi1997continuous,carr2010understanding,vojta2000quantum,carollo2020geometry}. While this framework has been highly successful, it fails to describe a broad class of quantum phases and transitions where no symmetry is broken, such as localization transitions~\cite{anderson1958absence,abrahams1979scaling,segev2013anderson,lee1981anderson,giordano1979experimental,jeon2023localization,jeon2025electron}, topological phases~\cite{basu2023topological,senthil2015symmetry,moessner2021topological}, and critical states with strongly inhomogeneous wave functions~\cite{ryu1992extended,kohmoto1983metal,kohmoto1986electronic,jeon2021topological}. This calls for alternative, symmetry-independent probes that directly reflect the structure of the quantum ground state, such as many-body entanglement~\cite{wu2004quantum,de2021entanglement,skinner2019measurement,google2023measurement,dutta2015quantum,osterloh2002scaling}, topological invariants~\cite{nakahara2018geometry,bernevig2013topological,yang2019topology}, and localization length~\cite{vosk2015theory,shima2004localization,de1998delocalization,jeon2026delocalization,wang2025quantum}.

One promising route is provided by a quantum geometric perspective on many-body states~\cite{carollo2020geometry,jiang2025revealing,nakahara2018geometry,yu2025quantum,liu2025quantum,de2023mapping,kang2025measurements,gianfrate2020measurement}. In this view, quantum phases are distinguished by the geometric structure of their ground-state wave functions, where the notion of metric between quantum states, such as quantum weight plays a central role~\cite{onishi2025quantum,onishi2024topological,onishi2024fundamental,ghosh2024probing,gianfrate2020measurement}. This idea has been extensively studied in periodic systems with well-defined band structures—most notably flat-band systems—where quantum geometry controls physical responses beyond symmetry-based classifications, while it remains underexplored in aperiodic and disordered systems~\cite{tian2023evidence,herzog2022superfluid,julku2021quantum,kruchkov2022quantum,mera2022nontrivial}. 

Recently, the concept of hyperuniformity that characterizes states in which long-wavelength density fluctuations are anomalously suppressed
has opened a new avenue for distinguishing phases of matter~\cite{torquato2003local,torquato2018hyperuniform,torquato2016hyperuniformity,torquato2021local,wan2023hyperuniform,leseur2016high,gkantzounis2017hyperuniform,aubry2020experimental,huang2021circular,hexner2017enhanced}.
This 
framework has been successfully applied to a wide range of disordered and quasiperiodic systems, providing 
a way to characterize spatially inhomogeneous distributions in these systems~\cite{torquato2018hyperuniform,torquato2016hyperuniformity,oguz2016hyperuniformity,lin2017hyperuniformity,chen2023disordered,sgrignuoli2022subdiffusive,fuchs2019landau,crowley2019quantum,torquato2021structural,sakai2022hyperuniform,sakai2022quantum,koga2024hyperuniform,hori2024multifractality,koga2025critical,vanoni2025effective,wang2025hyperuniform,jeon2026delocalization,chen2025anomalous}.
However, existing studies have primarily focused on fluctuations defined at a classical level, such as the variance of static density configurations or externally imposed potentials.
Hence, the role of intrinsic quantum fluctuations arising from many-body wave functions remains largely unexplored, despite their expected importance for quantum phase transitions.

In this Letter, we generalize the concept of hyperuniformity to incorporate quantum effects, 
proposing the quantum hyperuniformity, defined by the suppression of long-wavelength 
quantum fluctuations in a many-body state. 
We show that the charge-density distribution
at zero temperature 
is generally quantum hyperuniform (QHU), reflecting the charge conservation. Nevertheless, we find that the 
degree of the uniformity, defined by the scaling exponent of long-wavelength density structure factor can be classified 
into several different classes, depending on the nature of the many-body state.
Remarkably, these classes distinguish not only conventional gapped and gapless phases but also critical phases in which the wave functions are multifractal.
Thus, the 
QHU class
serves as a direct fingerprint of quantum criticality. Notably, the 
QHU
classification of the charge density acts in a complementary manner to its classical counterpart, enabling a systematic characterization of the intricate gap structure of quasiperiodic systems.
Furthermore, we show that the scaling of the quantum weight~\cite{onishi2025quantum,onishi2024topological,onishi2024fundamental}—a quantum geometric quantity derived from the charge-density structure factor—provides direct access to the size of the underlying gap. Owing to the experimental accessibility of density structure factors, quantum hyperuniformity—together with its classical counterpart and the quantum weight—offers a practical and broadly applicable probe of quantum phase transitions, including those in quasiperiodic systems.

\textit{\tcb{Quantum hyperuniformity---}}
First, we 
briefly review the conventional hyperuniformity \cite{torquato2003local,torquato2016hyperuniformity,torquato2018hyperuniform} for a scalar field, which we call classical hyperuniformity in this Letter. Let us consider a scalar field, such as 
electron density and on-site potential
$\phi_i$, where $i$ is the site index.
Its spatial average is $\bar{\phi}=\sum_i\phi_i/N$, where $N$ is the system size. We consider the Fourier transformation of the fluctuation 
as $\delta\phi(\vec{q})=\sum_j(\phi_j-\bar{\phi})e^{-\mathrm{i}\vec{q}\cdot\vec{r}_j}$, where $\vec{r}_j$ is the position vector of site $j$. Then, the classical structure factor of 
$\phi_i$
is defined by
\begin{align}
    \label{classicalS}
    &S_C(\vec{q})=\frac{1}{N}\vert\delta\phi(\vec{q})\vert^2,
\end{align}
which indicates the fluctuation 
at a length scale 
proportional to $\vert\vec{q}\vert^{-1}$. In particular, for 
small $\vert\vec{q}\vert$,
$S_C(\vec{q})$
describes the fluctuation at a large length scale. If $\lim_{\vec{q}\to0}S_C(\vec{q})=0$, we call the scalar-field distribution $\{\phi_i\}$ classical hyperuniform (CHU), referring to the suppression of large-scale fluctuations. Note that $\lim_{\vec{q}\to0}S_C(\vec{q})>0$ for a random potential, which is nonhyperuniform. We further classify the strength of hyperuniformity in terms of the scaling power of $S_C(\vec{q})$ around $\vec{q}=0$. Specifically, when $S_C(\vec{q})\sim\vert\vec{q}\vert^\nu$, $\nu>1$, $\nu=1$ and $0<\nu<1$ define class I, II and III CHU, respectively~\cite{torquato2016hyperuniformity}. 
The CHU characterizes how spatial fluctuations of a classical quantity extend over long wavelength. Exploiting this property, recent studies have proposed a new perspective on phase transitions in aperiodic systems, viewing them as transitions between distinct CHU classes~\cite{sakai2022quantum} or as transitions from non-hyperuniform to hyperuniform states~\cite{hori2024multifractality}.

However, since 
$\phi_i$ is a classical variable, 
we cannot capture quantum fluctuations such as interference effect of wavefunctions and resulting non-local correlations~\cite{osterloh2002scaling,jeon2022length,jeon2025hidden,muller2022measurement,sachdev1999quantum,morr2006impurities}. To 
incorporate
such quantum effects, we generalize the definition of hyperuniformity 
by replacing 
a c-number $\phi_i$ with a q-number $\hat{\phi}_i$.
More explicitly, we incorporate the quantum fluctuation as $\delta\hat{\phi}(\vec{q})=\sum_j(\hat{\phi}_j-\bar{\phi})e^{-\mathrm{i}\vec{q}\cdot\vec{r}_j}$, where $\bar{\phi}=\sum_i\langle\hat{\phi}_i\rangle/N$ and $\langle \rangle$ means the thermal average of many-body states (ground-state expectation value at zero temperature), and define the structure factor 
by
\begin{align}
    \label{structureT}
    &S(\vec{q})=\frac{1}{N}\langle\vert  \delta\hat{\phi}(\vec{q})  \vert^2\rangle.
\end{align}
In x-ray scattering experiments, we generally obtain a signal from the structure factor in Eq.~\eqref{structureT} that mixes classical and quantum fluctuation effects~\cite{macchi2013modern,schulke1995dynamic,balut2025quantum,cohen2016fundamentals}. From now on, let us focus on the quantum contribution of the structure factor, $S_Q(\vec{q})=S(\vec{q})-S_C(\vec{q})$. Since $\langle \hat{\phi}_j-\langle\hat{\phi_j}\rangle\rangle=0$, this reads \begin{align}
    \label{structureQ}
    &S_Q(\vec{q})=\frac{1}{N}\langle\vert  \delta_Q\hat{\phi}(\vec{q})  \vert^2\rangle,
\end{align}
where $\delta_Q\hat{\phi}(\vec{q})=\delta\hat{\phi}(\vec{q})-\delta\phi(\vec{q})=\sum_j(\hat{\phi}_j-\phi_j)e^{-\mathrm{i}\vec{q}\cdot\vec{r}_j}$ and $\phi_j=\langle\hat{\phi}_j\rangle$. We call the system QHU if  $\lim_{\vec{q}\to 0} S_Q(\vec{q})=0$.
We emphasize that 
$S_Q(\vec{q})$
represents the quantum fluctuation, including thermal fluctuations at finite temperatures, rather than classical variance encoded in Eq.~\eqref{classicalS}. At finite temperatures, $\lim_{\vec{q}\to0}S_Q(\vec{q})>0$, i.e., non-QHU while at zero temperature, the thermal average in Eq.~\eqref{structureQ} reduces to the expectation value for the many-body ground state, implying that $\lim_{\vec{q}\to0}S_Q(\vec q)=0$ if $\sum_i \langle\hat{\phi}_i\rangle$ is conserved. Nevertheless, the scaling behaviors in the  small-$\vert\vec{q}\vert$ limit are diverse, depending on the nature of the many-body ground state, as we will show below. 

\begin{figure}[h]
    \centering
    \includegraphics[width=0.5\textwidth]{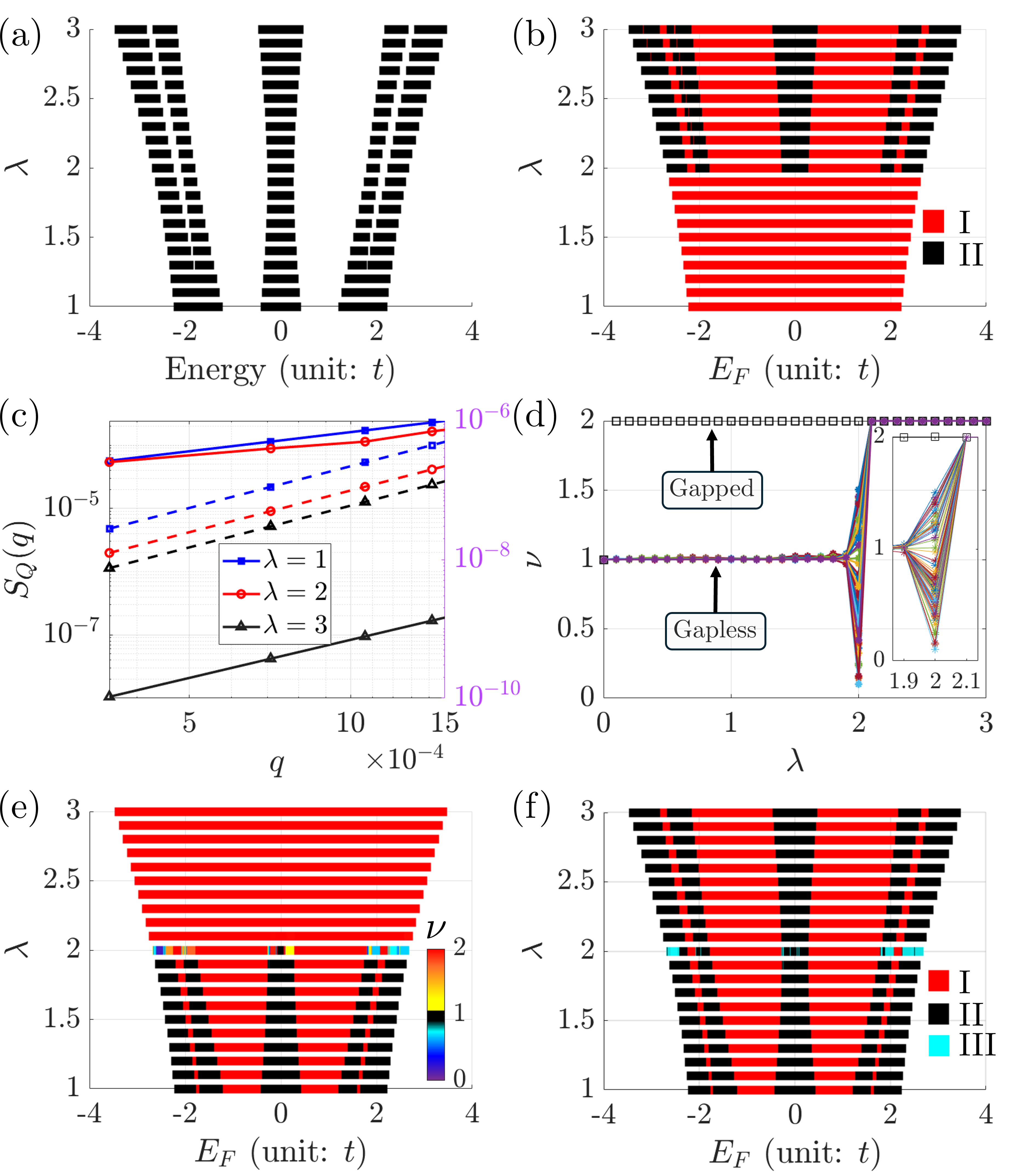}
    \caption{(a) Spectrum and (b) phase diagram of CHU class for the Aubry–André model. Note that tiny gaps are present but invisible in this resolution. See Supplemental Material \cite{supplementQWAA} for a more detailed structure. (c) Scaling behaviors of $S_Q$ for various $\lambda$ values. Solid lines (left axis) indicate the results for a Fermi level crossing a band,
    of which the scaling exponents are $1$, $0.618$, and $2$ for $\lambda=1,2$, and $3$, respectively.
    Dashed lines (right axis) indicate the results for a Fermi level inside a gap, of which the scaling exponent is $2$ for all $\lambda$. (d) Scaling exponent, $\nu$, plotted against $\lambda$. The black open square indicates the results for the Fermi level inside a gap, while the other symbols represent those for distinct Fermi levels crossing a band. The inset is drawn to emphasize a diversification of $\nu$ at $\lambda=2$. 
    (e) The landscape of $\nu$ plotted against $\lambda$ and the Fermi energy $E_F$. (f) Phase diagram of total hyperuniformity classes incorporating classical and quantum fluctuations. The red, black and skyblue colors represent class I, II and III, respectively. The system size is $N=17711$.
    }
    \label{fig: 1}
\end{figure}

\textit{\tcb{Quantum hyperuniform classification---}} We now apply above QHU classification 
to quantum phase transition systems. As a concrete example, let us consider the Aubry–André model, which is an extensively studied one-dimensional (1D) quasiperiodic model showing a localization-delocalization transition at a finite strength of the quasiperiodic potential~\cite{aubry1980analyticity,bu2022quantum,hetenyi2025numerical,sinha2019kibble,biddle2011localization,an2021interactions}. We emphasize that our results can be extended to systems exhibiting localization–delocalization transitions with mobility edges~\cite{supplementQWAA,miranda2024mechanical}.
The Aubry–André Hamiltonian 
reads
\begin{align}
    \label{H}
    &H=-t\sum_{i=1}^N (c_i^\dagger
c_{i+1}+c_{i+1}^\dagger c_i)+\lambda\sum_{i=1}^N \cos(2\pi\varphi i)c_i^\dagger c_i.
\end{align}
Here, $c_i (c_i^\dagger)$ is the annihilation (creation) operator of a spinless fermion at 
site $i$, and $t$ is the uniform hopping magnitude. We set $t=1$ as the energy unit. To avoid the boundary effect, we consider periodic boundary condition, $c_{N+1}=c_1$ for the system size $N=F_n$. Here, $F_n$ is the $n$-th Fibonacci number. $\varphi=F_{n-1}/F_n$ is a rational approximant of the golden ratio, $(\sqrt{5}-1)/2$. Figure~\ref{fig: 1}(a) exhibits the spectrum of the Aubry–André model. The localization characteristics of single-particle wavefunction of this model has been widely studied~\cite{aubry1980analyticity,bu2022quantum,hetenyi2025numerical,sinha2019kibble,biddle2011localization,an2021interactions}. Notably, the Hamiltonian in Eq.~\eqref{H} admits the self-duality at $\lambda=2$, and hence the 
wavefunctions are localized (extended) for $\lambda>(<) 2$~\cite{aubry1980analyticity}. At the quantum critical point, $\lambda=2$, multifractal critical states, which are neither localized nor extended but show a power-law scaling, emerge in the whole spectrum~\cite{aubry1980analyticity,bu2022quantum,hetenyi2025numerical,sinha2019kibble,biddle2011localization,an2021interactions}. At zero temperature, the many-body ground state exhibits CHU charge-density distribution for general Fermi levels. 
However,
the CHU class of charge density, $n_i=\langle c_i^\dagger c_i\rangle$, changes as a function of $\lambda$ and Fermi energy: 
It changes from I to II within the spectrum as $\lambda$ exceeds the critical value, $2$, while it remains class I when the Fermi energy lies within a gap. Hence, for $\lambda<2$, the charge density belongs to CHU class I, whereas for $\lambda\ge 2$ it belongs to either CHU class I or II, depending on whether the Fermi energy lies within the gap [see Fig.~\ref{fig: 1}(b)]~\cite{sakai2022quantum}.

In this Letter, we take account of the quantum effect, applying the QHU concept to the Aubry–André model.
Let us consider the charge-density operator, $\hat{n}_i=c_i^\dagger c_i$ as an observable. Then, the structure factor of Eq.~\eqref{structureQ} becomes
\begin{align}
    \label{Scharge}
    &S_Q(q)=\frac{1}{N}\sum_{i,j}e^{-\mathrm{i}q(x_i-x_j)}(\langle \hat{n}_i\hat{n}_j \rangle-\langle\hat{n}_i\rangle\langle\hat{n}_j\rangle).
\end{align}
$C_{ij}\equiv\langle \hat{n}_i\hat{n}_j \rangle-\langle\hat{n}_i\rangle\langle\hat{n}_j\rangle$ is known as a connected correlation matrix. In non-interacting systems, we can apply Wick's theorem, so that $C_{ij}$ becomes
\begin{align}
    \label{connectedcorr}
    &C_{ij}=\langle c_i^\dagger c_j\rangle(\delta_{ij}-\langle c_j^\dagger c_i\rangle).
\end{align}
Since the system obeys $U(1)$ symmetry, 
$S_Q(q\to 0)$ vanishes, i.e.,
QHU at zero temperature.

Before 
going into the calculated results,
we briefly summarize the scaling behavior of 
$S_Q(q)$
in ordinary cases. First, when the ground state is gapped, we have $C_{ij}\sim e^{-\vert i-j\vert/\xi}$ with the effective localization length $\xi$, which is inversely proportional to the gap size. In this case, $S_Q(q)=Kq^2/2+\mathcal{O}(q^4)$
for small $|q|$. Here, the constant $K=\frac{\partial^2S_Q}{\partial q^2}(q=0)$ is called the quantum weight, which measures the localization length of the Wannier function~\cite{onishi2025quantum}. We will show below that 
$K$
would be used to 
estimate the gap size. Second, 
for a 1D gapless periodic system, $C_{ij}\sim \vert i-j\vert^{-1}$
and hence $S_Q(q)\propto \vert q\vert$~\cite{supplementQWAA,cohen2016fundamentals}
.
Thus, while these cases are both QHU, the QHU class changes from I to II as either the gap is closed or delocalization transition occurs.
Namely, the QHU class of charge density can probe the quantum phase transition between gapped and gapless phases without resorting to a calculation of the spectrum.

For $\lambda>0$ in the Aubry–André model, infinitely many spectral gaps exist in the spectrum. These gaps are labeled by a topological index known as the gap label~\cite{kellendonk2015mathematics}. In detail, we have a spectral gap when the filling fraction is $a\varphi+b$ with integers $a,b$ known as the gap labels~\cite{kellendonk2015mathematics,kaminker2003proof,damanik2023gap,hetenyi2025numerical}. Based on these gap labels, we consider separately the cases in which the Fermi level lies within these gaps and those in which it does not. We investigate the scaling power $\nu$ of 
$S_Q(q)\sim q^\nu$ 
for a small $q$ 
as a function of $\lambda$ for various Fermi levels.

Figure~\ref{fig: 1}(c) shows typical small-$q$ behaviors of $S_Q(q)$ for different $\lambda$ values. We note that the scaling exponent $\nu$ of $S_Q(q\sim 0)$ depends on $\lambda$ when the Fermi level $E_F$ is not in the gap. In detail, $\nu=1 (2)$ for $\lambda< (>)2$, reflecting the extended and localized phases, respectively. When 
$E_F$
lies inside a gap, 
$\nu=2$
for all $\lambda>0$. Interestingly, we observe a pronounced diversification of 
$\nu$ from 0 to 2
at the quantum critical point, $\lambda=2$ [see the inset of Fig.~\ref{fig: 1}(d)],
in particular, $0<\nu<1$ for some filling fractions, indicating class-III QHU. This is because the multifractal critical states in quasiperiodic systems enable unconventional long-range interstate density correlations. To be more specific, we rewrite Eq.~\eqref{Scharge} in terms of single-particle eigenstates of the Hamiltonian,
\begin{align}
    \label{Squantum_occ}
    &S_Q(q)=\sum_{n\in occ}\sum_{m\in unocc}\vert \langle n\vert e^{\mathrm{i}q\hat{x}}\vert m\rangle \vert^2,
\end{align}
where $n,m$ are the indices of the eigenstates, and $occ (unocc)$ denotes occupied (unoccupied) states. $\hat{x}$ is the position operator~\cite{supplementQWAA}. Thus, the small-$q$ behavior of $S_Q(q)$ originates from the extent to which an occupied-state momentum-space wavefunction overlaps with unoccupied states when a small translation in momentum space is applied. At $\lambda=2$, all single-particle wavefunctions are critical and multifractal, being neither localized nor extended in both real and momentum spaces, with the momentum-space wavefunction exhibiting fractal mixing of multiple momentum components~\cite{liu2024dual,supplementQWAA}. Hence, the wavefunction overlap under small-momentum translations would be anomalously enhanced compared to larger momentum shifts, giving rise to a concave small-$q$ structure factor for certain filling fractions. Note that such a class-III QHU structure factor is a characteristic feature of multifractal states in general quasiperiodic systems~\cite{supplementQWAA}.

The scaling exponent $\nu$ characterizes the decay of long-range charge correlations, with smaller 
$\nu$ indicating a slower decay of $C_{ij}$. Notably, QHU class I (III) shows a faster (slower) decay of averaged long-range correlations compared to conventional 1D gapless system, $C_{ij}\sim\vert i-j\vert^{-1}$~\cite{supplementQWAA}.


Figure~\ref{fig: 1}(e) illustrates the landscape of the exponent $\nu$ as a function of $\lambda$ and $E_F$. 
Note that the QHU class I appears for all $\lambda>0$. Specifically, for $\lambda>2$, where the states are localized, arbitrary filling fractions correspond to the QHU class I, while for $0<\lambda\le2$, the QHU class I appears only for the filling fraction $a\varphi+b$ (with integers $a$ and $b$) of gap labels. This is attributed to the suppression of quantum fluctuation between occupied and unoccupied states by localized states or a gapped phase. Whereas, QHU class II describes band regions with finite density of states for $\lambda\leq 2$, for which gapless excitations dominate and the structure factor exhibits enhanced infrared fluctuations. 
Notably, QHU class III behavior can also emerge precisely at the critical point $\lambda=2$, where the multifractal critical wavefunctions appear
and the scaling exponent $\nu$ takes anomalous values  
other than $1$ and $2$.
We summarize the QHU classification in Table~\ref{table:summary}.
\begin{table}[t]
\centering
\caption{QHU classification of charge fluctuations in Aubry–André model. $\varphi=\frac{\sqrt{5}-1}{2}$.\label{table:summary}}
\begin{tabular}{c c c}
\hline\hline
QHU & $\lambda$ & Filling fraction \\
\hline
Class I   & $0<\lambda\le2t$ 
          & $\in\mathbb{Z}+\varphi\mathbb{Z}$\\ & $\lambda>2t$& arbitrary\\
Class II  & $\lambda \le 2t$ 
          & $\notin\mathbb{Z}+\varphi\mathbb{Z}$ \\
Class III & $\lambda = 2t$ 
          & $\notin\mathbb{Z}+\varphi\mathbb{Z}$ \\
\hline\hline
\end{tabular}
\label{tab:alpha_classification}
\end{table}

Comparing Figs.~\ref{fig: 1}(b) and ~\ref{fig: 1}(e), we emphasize that the QHU and CHU classes play complementary roles in distinguishing phases in the Aubry–André model. Note that the parameter regimes in which class II emerges in the gapless phase are opposite for the CHU and QHU. Specifically, CHU class II emerges only for $\lambda\ge2$, while QHU class II emerges only for $\lambda\le2$. Thus, the positions of spectral gaps are completely characterized by using both CHU and QHU classes.
This complementarity stems from the different fluctuations probed by classical and quantum contributions of the structure factor. $S_C$ measures how much the local expectation value of the charge density deviates from its spatial average. Hence, in regimes where wave functions have strong spatial inhomogeneity—namely localized or critical phases ($\lambda\ge2)$—CHU distinguishes gapped and gapless phases through a 
behavior
of class I and II, respectively. While in the extended regime ($\lambda<2$), where wavefunctions spread over the entire system, CHU remains class I regardless of the presence or absence of a gap.
In contrast, $S_Q$ probes quantum fluctuations arising from particle–hole excitations near the Fermi level. Consequently, QHU is sensitive to the presence or absence of a gap in the extended and critical regimes ($\lambda\le2$), where wavefunction overlap is significant, while in the localized regime ($\lambda>2$), where such overlap is negligible, it exhibits class I behavior independent of the gap. Moreover, at the critical point $\lambda=2$, the exotic spatial structure of multifractal wavefunctions allows for the emergence of QHU class III. Thus, CHU and QHU classes identify spectral gaps in localized and extended regimes, respectively, without explicit spectral calculations.

Experimentally, one measures the structure factor in Eq.~\eqref{structureT}, which includes both classical and quantum contributionss~\cite{schulke1995dynamic,balut2025quantum,cohen2016fundamentals,chang2012direct,frano2016long,teng2022discovery,macchi2013modern,ghosh2024probing}. Figure~\ref{fig: 1}(f) illustrates a 
phase diagram of the hyperuniformity classification 
based on Eq.~\eqref{structureT}, 
merging the CHU and QHU classifications. 
This total
hyperuniformity is governed by the smaller scaling exponent $\nu$ of CHU and QHU. Specifically, for all $\lambda$ values, class I (II or III) hyperuniformity appears in the gapped (gapless) phase. Notably, for $\lambda = 2$, the class-III QHU behavior is not masked by class-II CHU.
Thus, the total hyperuniformity not only distinguishes gapped and gapless phases for all $\lambda$, but also detects charge density correlations that decay anomalously slowly at long distances due to critical states.

\begin{figure}[h]
    \centering
    \includegraphics[width=0.5\textwidth]{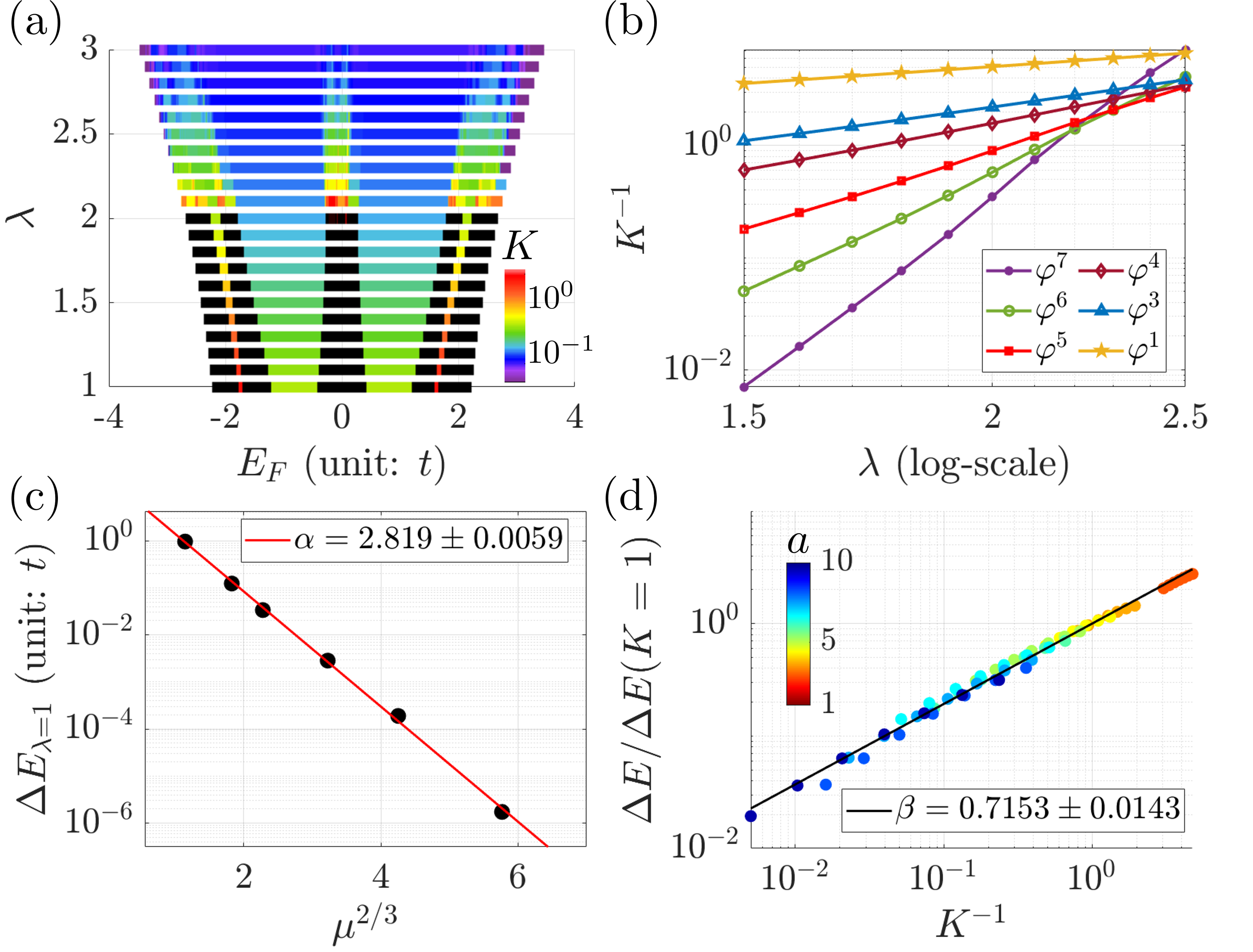}
    \caption{(a) The landscape of quantum weight $K$ plotted against $\lambda$ and $E_F$. Black indicates infinite $K$. (b) Log--log plot of $K^{-1}$ for the filling fractions, $\varphi^{s}=F_s\varphi+F_{s-1}$ with integers $s\ge 1$, which correspond to the gaps of the Aubry–André model, as a function of $\lambda$. These quantum weights exhibit scaling behavior, $K\propto\vert \lambda\vert^{-\mu}$. (c) The gap size $\Delta E$ at $\lambda=1$ for different gaps as a function of the scaling exponent $\mu$, which show a relation $\Delta E_{\lambda=1} \propto e^{-\alpha\mu^{2/3}}$. (d) Gap sizes at different filling fractions as functions of $K^{-1}$, normalized by their values at $K=1$, which show $K^{-\beta}$ universal scaling. Each color represents a gap for $1.5\le\lambda\le2$ at a filling fraction $a\varphi + b$ with a different integer $a$. The system size is $N=17711$.
    }
    \label{fig: 2}
\end{figure}

\textit{\tcb{Gap size and quantum weight---}} Motivated by this complementary role of CHU and QHU classes as the probes of gaps in the localized and extended regime, respectively, we now show that the quantum weight can be used to characterize the size of gaps in the delocalized regime ($\lambda\le2$). Note that a class-I QHU structure factor,
$S_Q(q)=K q^{2}/2+\mathcal{O}(q^{4})$, with $K$ the quantum weight,
generically arises when the Fermi level lies inside a gap, regardless of the values of $\lambda\neq0$ [see Fig.~\ref{fig: 2}(a)].

Figure~\ref{fig: 2}(b) 
plots $K^{-1}$ against
$\lambda$ for various Fermi levels inside different gaps. Specifically, we consider the filling fractions of the form $\varphi^{s}=F_s\varphi+F_{s-1}$ with integer $s\ge1$, which correspond to the gaps of the Aubry–André model. The quantum weight generally decreases as $\lambda$ increases since it measures the localization length of the Wannier function. Hence, we have universal scaling behavior given by $K\propto\vert\lambda\vert^{-\mu}$, where $\mu$ is an exponent 
which depends
on 
the gap label.
Note that the exponent $\mu$ represents the sensitivity of the localization strength 
to $\lambda$. 
We find that for $E_F$ inside a gap, the exponent $\mu$ reflects the size of the gap. 
Since a larger gap more strongly suppresses virtual particle-hole excitations, the quantum weight for a larger gap is less sensitive to variations of $\lambda$. We observe an universal scaling relation between $\mu$ and the gap size at $\lambda=1$, $\Delta E_{\lambda=1}\propto e^{-\alpha\mu^{2/3}}$ with $\alpha=2.8190\pm0.0059$ [see Fig.~\ref{fig: 2}(c)]. Thus, the scaling of the quantum weight enables an estimation of the gap size around the Fermi level.

Surprisingly, the sizes of different gaps follow a universal power-law scaling as a function of the quantum weight for $\lambda\le2$, i.e.,  $\Delta E\propto K^{-\beta}$ with $\beta=0.7153\pm0.0143$ in the Aubry–André model [see Fig.~\ref{fig: 2}(d)]. This reflects that a wider gap admits a smaller localization length for $\lambda\le2$. On the other hand, in the localized regime, $\lambda>2$, the gap size can 
increase with $K$,
depending on the gap position, because the localization length is no longer solely determined by the gap size~\cite{supplementQWAA}.




\textit{\tcb{Conclusion---}} In conclusion, we presented quantum hyperuniformity as a powerful probe of localization properties in many-body ground states. While charge fluctuations at zero temperature are generally suppressed by particle-number conservation, their infrared scaling defines distinct quantum-hyperuniform classes that sharply discriminate gapped, gapless, localized, and extended phases. In the Aubry–André model, quantum hyperuniformity directly captures the localization transition: gapped phases exhibit quadratic scaling of the structure factor, gapless extended phases show linear behavior, and at the self-dual critical point multifractal wave functions induce anomalous concave infrared scaling, reflecting enhanced particle-hole excitation probability and anomalously slowly decaying charge-density correlation. It turns out that classical and quantum hyperuniformities distinguish gapped and gapless phases in localized and extended regimes, respectively. Since both contributions coexist in experimentally measured structure factor, the total structure factor enables the identification of not only gapped and gapless phases but also anomalously slowly decaying long-range charge-density correlations arising from critical states. Finally, the quantum weight associated with quadratic scaling serves as a quantitative measure of gap sizes, enabling spectral information to be extracted solely from ground-state correlations. Given the direct experimental accessibility of density structure factors and quantum weight~\cite{kang2025measurements,ghosh2024probing,gianfrate2020measurement}, our results establish quantum hyperuniformity as a broadly applicable and experimentally relevant fingerprint of quantum criticality and long-wavelength quantum fluctuations.

\section*{Acknowledgments}
This work was supported by JSPS KAKENHI Grant No. JP25H01397 and JP25H01398.

\bibliography{reference}

\newpage

\begin{widetext}

\renewcommand{\thefigure}{S\arabic{figure}}
\renewcommand{\theequation}{S\arabic{equation}}

\title{Supplemental Materials for\\ ``Quantum Hyperuniformity and Quantum Weight"
}

\author{Junmo Jeon}
\email{junmojeon@sophia.ac.jp}
\affiliation{Physics Division, Sophia University, Chiyoda-ku, Tokyo 102-8554, Japan}
\author{Shiro Sakai}
\email{shirosakai@sophia.ac.jp}
\affiliation{Physics Division, Sophia University, Chiyoda-ku, Tokyo 102-8554, Japan}

\maketitle

\section{Derivation of alternative form of structure factor}

In this section, we derive an alternative form of quantum contribution of the charge-density structure factor in terms of wave functions overlap. We consider a noninteracting fermionic system whose many-body ground state is a
Slater determinant constructed from single-particle eigenstates.

Let us define the local number operator at site $j$ as
\begin{align}
\hat{n}_j = c_j^\dagger c_j .
\end{align}
The connected density--density correlation matrix is then defined by
\begin{align}
C(j,j') =
\langle \hat{n}_j \hat{n}_{j'} \rangle
-
\langle \hat{n}_j \rangle \langle \hat{n}_{j'} \rangle .
\end{align}
Let $\psi_m(j)$ denote the single-particle eigenfunctions of the Hamiltonian.
The fermionic operators in real space can be expanded as
\begin{align}
c_j &= \sum_m \psi_m(j)\, c_m , \\
c_j^\dagger &= \sum_m \psi_m^*(j)\, c_m^\dagger .
\end{align}
Using this expansion, the local number operator becomes
\begin{align}
\hat{n}_j
=
\sum_{mn}
\psi_m^*(j)\psi_n(j)\,
c_m^\dagger c_n .
\end{align}
The product of two local number operators is given by
\begin{align}
\hat{n}_j \hat{n}_{j'}
=
\sum_{mn}
\sum_{pq}
\psi_m^*(j)\psi_n(j)
\psi_p^*(j')\psi_q(j')
\, c_m^\dagger c_n c_p^\dagger c_q .\label{eq:njnj}
\end{align}
Since the ground state is a Slater determinant, Wick's theorem applies exactly.
The relevant contraction reads
\begin{align}
\langle c_m^\dagger c_n c_p^\dagger c_q \rangle
=
\delta_{mn}\delta_{pq} \langle\hat{n}_m\rangle \langle\hat{n}_p
\rangle+
\delta_{mq}\delta_{np} \langle\hat{n}_m \rangle(1-\langle\hat{n}_n\rangle),\label{eq:wick}
\end{align}
where $\langle\hat{n}_m\rangle$ is the occupation number of the single-particle eigenstate $m$. Note that at zero temperature, $\langle\hat{n}_m\rangle\in\{0,1\}$. When we take the ground-state average of Eq.~(\ref{eq:njnj}), 
the first term on the r.h.s. of Eq.~(\ref{eq:wick}) reproduces
$\langle \hat{n}_j\rangle \langle \hat{n}_{j'}\rangle$ and therefore cancels in the connected
correlation function.
The remaining contribution yields
\begin{align}
C(j,j')
=
\sum_{mn}
\langle\hat{n}_m\rangle(1-\langle\hat{n}_n\rangle)\,
\psi_m^*(j)\psi_n(j)
\psi_n^*(j')\psi_m(j') .
\end{align}

The charge structure factor is defined as the Fourier transform of the connected
correlation matrix,
\begin{align}
S_Q(q)
=\frac{1}{N}
\sum_{j,j'}
e^{-\mathrm{i}q(x_j-x_{j'})}
\, C(j,j') .
\end{align}
Substituting the expression for $C(j,j')$, we obtain
\begin{align}
S_Q(q)
=\frac{1}{N}
\sum_{mn}
\langle\hat{n}_m\rangle(1-\langle\hat{n}_n\rangle)
\sum_{j,j'}
e^{-\mathrm{i}q(x_j-x_{j'})}
\psi_m^*(j)\psi_n(j)
\psi_n^*(j')\psi_m(j') .
\end{align}
Separating the sums over $j$ and $j'$, the structure factor can be written as
\begin{align}
S_Q(q)
=\frac{1}{N}
\sum_{mn}
\langle\hat{n}_m\rangle(1-\langle\hat{n}_n\rangle)
\left(
\sum_j e^{-\mathrm{i}qx_j}\psi_m^*(j)\psi_n(j)
\right)
\left(
\sum_{j'} e^{\mathrm{i}qx_{j'}}\psi_n^*(j')\psi_m(j')
\right) .
\end{align}
This leads to the compact expression
\begin{align}
S_Q(q)
=\frac{1}{N}
\sum_{mn}
\langle\hat{n}_m\rangle(1-\langle\hat{n}_n\rangle)
\left|
\sum_j e^{-\mathrm{i}qx_j}\,
\psi_m^*(j)\psi_n(j)
\right|^2 .
\end{align}

Finally, since $\langle\hat{n}_m\rangle(1-\langle\hat{n}_n\rangle)$ is nonzero only when $m$ is occupied and $n$ is unoccupied
in the ground state, the structure factor can be written as
\begin{align}
\label{mainresult}
S_Q(q)
=\frac{1}{N}
\sum_{m\in\mathrm{occ}}
\sum_{n\in\mathrm{unocc}}
\left|
\sum_j e^{-\mathrm{i}qx_j}\,
\psi_m^*(j)\psi_n(j)
\right|^2 .
\end{align}
Note that no assumption of translational symmetry is made, so the derivation applies equally
to periodic, quasiperiodic, and disordered systems.

\section{Scaling behaviors of structure factor in periodic systems}
\label{sec:-1}

\subsection*{Extended states and linear infrared scaling of the structure factor}

We first consider a one-dimensional periodic lattice system of length $N$ with translational invariance and spinless fermions at zero temperature. The Fourier transform of local charge density operator is 
\begin{align}
\hat{n}_q = \sum_{j=1}^N e^{-\mathrm{i}qx_j} \hat{n}_j ,
\qquad q = \frac{2\pi m}{N}.
\end{align}
Here, $0\le m\le N-1$ is an integer. The quantum contribution of static charge density structure factor is defined as
\begin{align}
S_Q(q) = \frac{1}{N}
\left(
\langle \hat{n}_{-q} \hat{n}_q \rangle
- \langle \hat{n}_{-q} \rangle \langle \hat{n}_q \rangle
\right).
\end{align}
Translational invariance implies $\langle \hat{n}_q \rangle = 0$ for all $q \neq 0$.

For a periodic system with extended single-particle eigenstates, the many-body ground state is a Slater determinant of Bloch states. Writing the density operator in momentum space,
\begin{align}
\hat{n}_q = \sum_k c_{k+q}^\dagger c_k ,
\end{align}
the structure factor can be expressed as
\begin{align}
S_Q(q) = \frac{1}{N} \sum_{k,k'}
\langle c_k^\dagger c_{k+q}
c_{k'+q}^\dagger c_{k'} \rangle .
\end{align}
Since the ground state is Gaussian, Wick's theorem applies and yields
\begin{align}
S_Q(q) = \frac{1}{N} \sum_k n_k (1 - n_{k+q}),
\end{align}
where the momentum occupation number for a Fermi surface ground state at zero temperature is
\begin{align}
n_k =
\begin{cases}
1, & |k| < k_F , \\
0, & |k| > k_F .
\end{cases}
\end{align}

For small but finite momentum $q$, the factor $n_k (1 - n_{k+q})$ is nonzero only when $k$ lies inside the Fermi sea while $k+q$ lies outside. This restricts $k$ to momentum intervals of width $|q|$ near the Fermi points $\pm k_F$. Consequently, the number of contributing momentum states scales linearly with $|q|$. In the thermodynamic limit, the sum can be replaced by an integral,
\begin{align}
S_Q(q\le k_F) = \int \frac{dk}{2\pi} \,
n_k (1 - n_{k+q})
= \frac{|q|}{2\pi}.
\end{align}
Thus, for periodic systems with extended and gapless states, the charge density structure factor exhibits a linear infrared scaling,
\begin{align}
S_Q(q) \propto |q| .
\end{align}

We emphasize that the linear infrared scaling of the charge density structure factor does not rely on translational invariance or Bloch momentum, but follows more generally from the presence of extended and gapless single-particle states. Recall that the static structure factor can be written as
\begin{align}
S_Q(q)
=
\frac{1}{N}
\sum_{e\neq 0}
\left|
\langle e | \hat{n}_q | 0 \rangle
\right|^2 ,
\label{eq:LehmannSq}
\end{align}
where $|0\rangle$ denotes the many-body ground state, $|e\rangle$ are excited many-body eigenstates, and
\begin{align}
\hat{n}_q = \sum_{j=1}^{N} e^{-\mathrm{i}q x_j} \hat{n}_j
\end{align}
is the Fourier component of the local density operator.

In systems with extended single-particle states, the density operator $\hat{n}_q$ couples the ground state to particle--hole excitations that are delocalized over the entire system. Crucially, in the absence of a spectral gap, such excitations can be generated with arbitrarily small energy cost as $q \to 0$. As a result, the dominant contribution to $S_Q(q)$ arises from low-energy particle--hole states with excitation energies $\omega_m = E_e - E_0 = O(q)$.

The number of such low-energy excitations is determined by the available phase space, which in one dimension grows linearly with the momentum transfer $|q|$. Moreover, the matrix elements $\langle e | \hat{n}_q | 0 \rangle$ remain finite in the thermodynamic limit for extended states, reflecting the fact that density fluctuations involve coherent contributions from the entire system rather than being confined to a finite region. Combining these observations, the sum in Eq.~\eqref{eq:LehmannSq} receives contributions from $O(|q|L)$ extended particle--hole states, each contributing a finite weight.

Consequently, in the long-wavelength limit one obtains
\begin{align}
S_Q(q) \propto |q| ,
\qquad q \to 0 ,
\end{align}
independent of whether translational symmetry is present. This establishes that linear infrared scaling of the charge density structure factor is a generic feature of extended, gapless states, rather than a consequence of crystalline order.

\subsection*{Gapped or localized phases}

For gapped phase, the many-body ground state is separated from excited states by a finite energy gap. Long-wavelength density fluctuations are therefore suppressed, and the static density response is analytic around $q=0$. Thus, the structure factor admits the expansion
\begin{align}
S_Q(q) = K q^2/2 + \mathcal{O}(q^4), \qquad q \to 0 .
\end{align}
The absence of a linear term reflects the lack of low-energy excitations at small momentum and serves as a characteristic signature of incompressible gapped phases.

Note that the correlation matrix of the localized phase is similar to the gapped phase, where the inverse of finite localization length plays the role of effective energy gap. Hence, the localized phases also generally exhibit quadratic scaling behavior of the structure factor.

\section{Various decaying behaviors of long-range charge correlations}
\begin{figure}[h]
    \centering
    \includegraphics[width=0.8\textwidth]{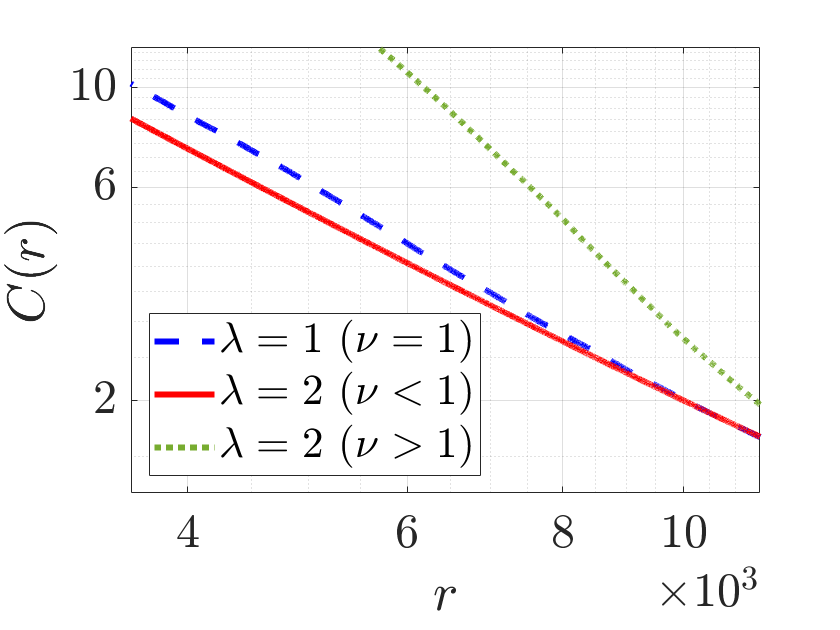}
    \caption{Log--log plot of the averaged long-range correlation function, $C(r)$ for different $\nu$ values. Here, the Fermi levels are not in the gap. Smaller $\nu$ corresponds to a slower decay of $C(r)$. The system size is $N=17711$. $t=1$.
    }
    \label{fig: 1supple}
\end{figure}
Figure~\ref{fig: 1supple} compares the distance dependence of the averaged correlation function,
\begin{align}
    \label{Cr}
    &C(r)=\frac{1}{Nr}\sum_{\vert i-j\vert\le r}C_{ij},
\end{align}
obtained from the connected correlation matrix $C_{ij}$ for different values of $\nu<2$.
For the gapless cases, QHU class II exhibits the conventional power-law decay $C(r)\sim 1/r$, while QHU class I (III) shows a faster (slower) power-law decay. Hence, the QHU classification shows that, in quasiperiodic systems, charge density correlations can exhibit anomalous power-law scaling induced by multifractal critical states.

\section{Gap size and quantum weight for localized regime}
In this section, we explore the sizes of gaps in terms of the quantum weight $K$ in the localized regime ($\lambda>2t$) of the Aubry–André model. Note that all wavefunctions are localized in this parameter regime. Figure~\ref{fig: supplegap} exhibits the sizes of gaps with different filling fractions as functions of $K$. Unlike the case of $\lambda\le2t$ discussed in the main text [Fig. 2(d)], the sizes of different gaps are not reduced to a universal scaling law. Moreover, we observe that the gap sizes for some filling fractions decrease with their $K$ values (see the blue curve in Fig.~\ref{fig: supplegap}, for instance). This indicates that the gap size no longer simply reflects the localization length of the Wannier function once the wavefunctions become localized. This observation is consistent with the nearly classical behavior of particles in the localized regime, where quantum effects are largely negligible.
\begin{figure}[h]
    \centering
    \includegraphics[width=0.8\textwidth]{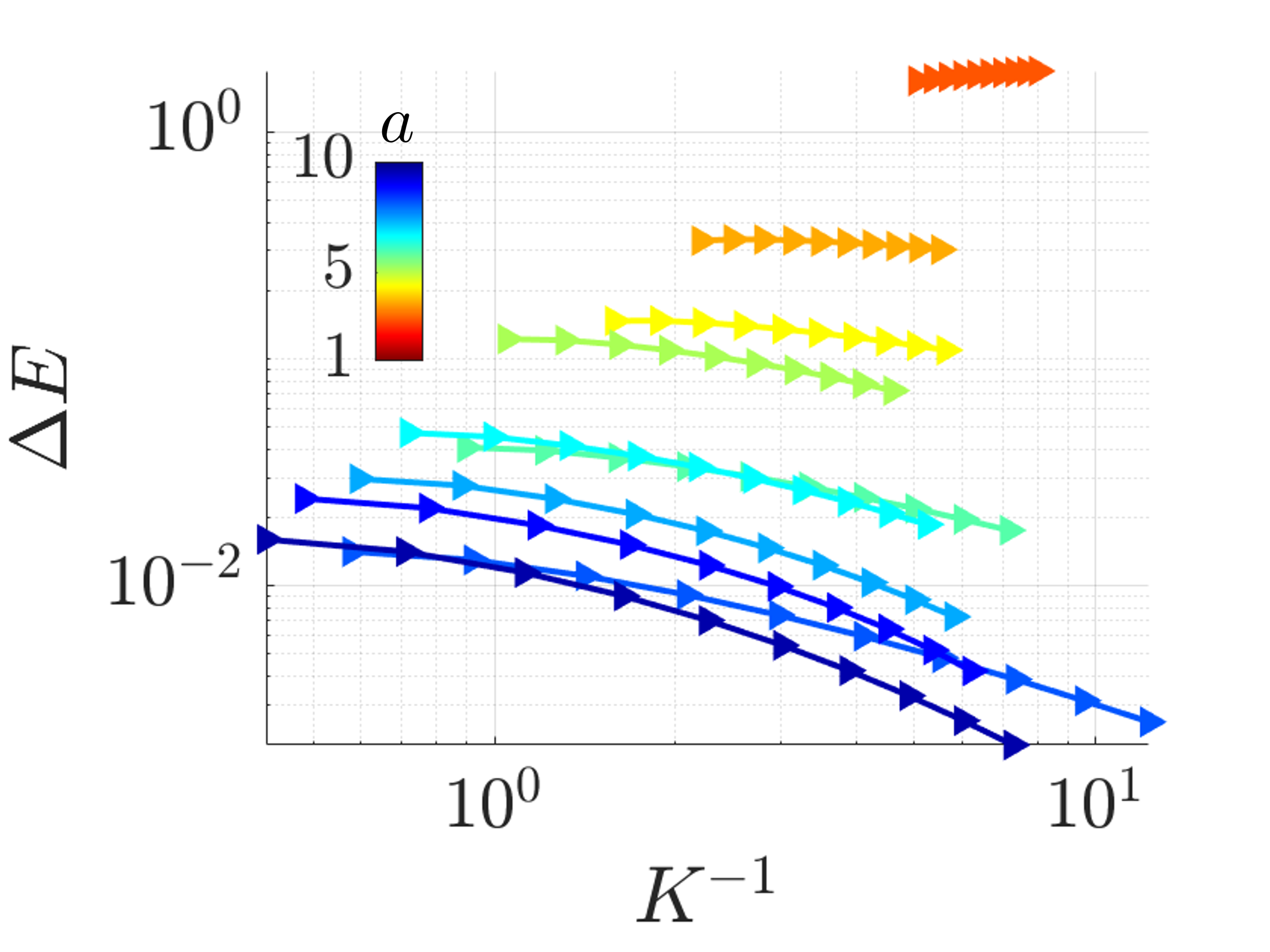}
    \caption{Gap sizes at different filling fractions as functions of $K$. Here, $\lambda> 2$. Each color represents a gap at a filling fraction $a\varphi + b$ with a different integer $a$. The system size is $N=17711$. $t=1$. Here, we use $\varphi=10946/17711$.
    }
    \label{fig: supplegap}
\end{figure}

\section{Quantum hyperuniformity in the presence of the mobility edges}
Our concept of quantum hyperuniformity (QHU) is not restricted to the Aubry--André (AA) model, but is broadly applicable to other quantum systems. 
Here, we focus on the QHU classification of a system with a mobility edge, realized by a modification of the AA model. 
The conventional AA model exhibits a critical point at $\lambda = 2t$ due to its self-duality, where all eigenstates are simultaneously extended, localized, or critical, and therefore does not host a mobility edge in its spectrum.

To introduce a mobility edge, we consider a generalized AA model with alternating hopping amplitudes $t_1$ and $t_2$~\cite{miranda2024mechanical}. 
\begin{align}
    \label{AASSH}
    &H=-\sum_{i=1}^N t(i)(c_i^\dagger
c_{i+1}+c_{i+1}^\dagger c_i)+\lambda\sum_{i=1}^N \cos(2\pi\varphi i)c_i^\dagger c_i.
\end{align}
Here, $t(i)=t_1$ if $\mathrm{mod}(i,2)=1$, and $t(i)=t_2$ for $\mathrm{mod}(i,2)=0$.
This model can be viewed as a hybrid of the Su--Schrieffer--Heeger (SSH) model and the AA model. We consider the periodic boundary condition.
For $t_1 \neq t_2$, the Hamiltonian supports both localized and extended eigenstates within the same spectrum, giving rise to a mobility edge. We set $t_1=1$ and $t_2=0.3$. We characterize the localization properties in terms of the inverse participation ratio (IPR) of the single-particle wavefunctions,
\begin{align}
&\mathrm{IPR}_n = \sum_{i} |\psi_n(i)|^4,
\end{align}
where $\psi_n(i)$ denotes the amplitude of the $n$th eigenstate at site $i$. 
From the system-size scaling of the IPR, $\mathrm{IPR}_n \sim N^{-D_f}$, we extract the fractal dimension $D_f$, which distinguishes extended ($D_f=1$), localized ($D_f=0$), and critical ($0<D_f<1$) states. We compute finite-size fractal dimension of $n$th eigenstate as
\begin{align}
    &D_f^{(n)}(N)=-\frac{\ln \mathrm{IPR}_n}{\ln N}.
\end{align}
Figure~\ref{fig: suppleAASSH}(a) presents a detailed landscape of the energy spectrum together with the localization properties of the eigenstates, in terms of finite-size fractal dimension. Note that the system has mobility edge for $\lambda\lesssim 1$.

By investigating the scaling behaviors of the structure factor, we find that the QHU class correctly identifies the delocalized states. In detail, the QHU class changes from II (III) to I when the Fermi level enters the gap if there are delocalized states [see $\lambda<1$ regime in Fig.~\ref{fig: suppleAASSH}(b)]. While, when the states are localized, the QHU class remains class-I QHU [see $\lambda>1$ regime in Fig.~\ref{fig: suppleAASSH}(b)]. Therefore, one can utilize the QHU classification to identify the mobility edge where the localization characteristic changes from delocalized to localized.

We emphasize that, in this model, the delocalization-to-localization transition driven by $\lambda$ occurs at energy-dependent critical values of $\lambda$. 
When the transition takes place at a given Fermi level, the associated critical state gives rise to class-III QHU. This shows that even in the presence of a mobility edge—where critical states exist only within a limited energy window—class-III QHU emerges precisely at the delocalization-to-localization transition point.

\begin{figure}[h]
    \centering
    \includegraphics[width=1\textwidth]{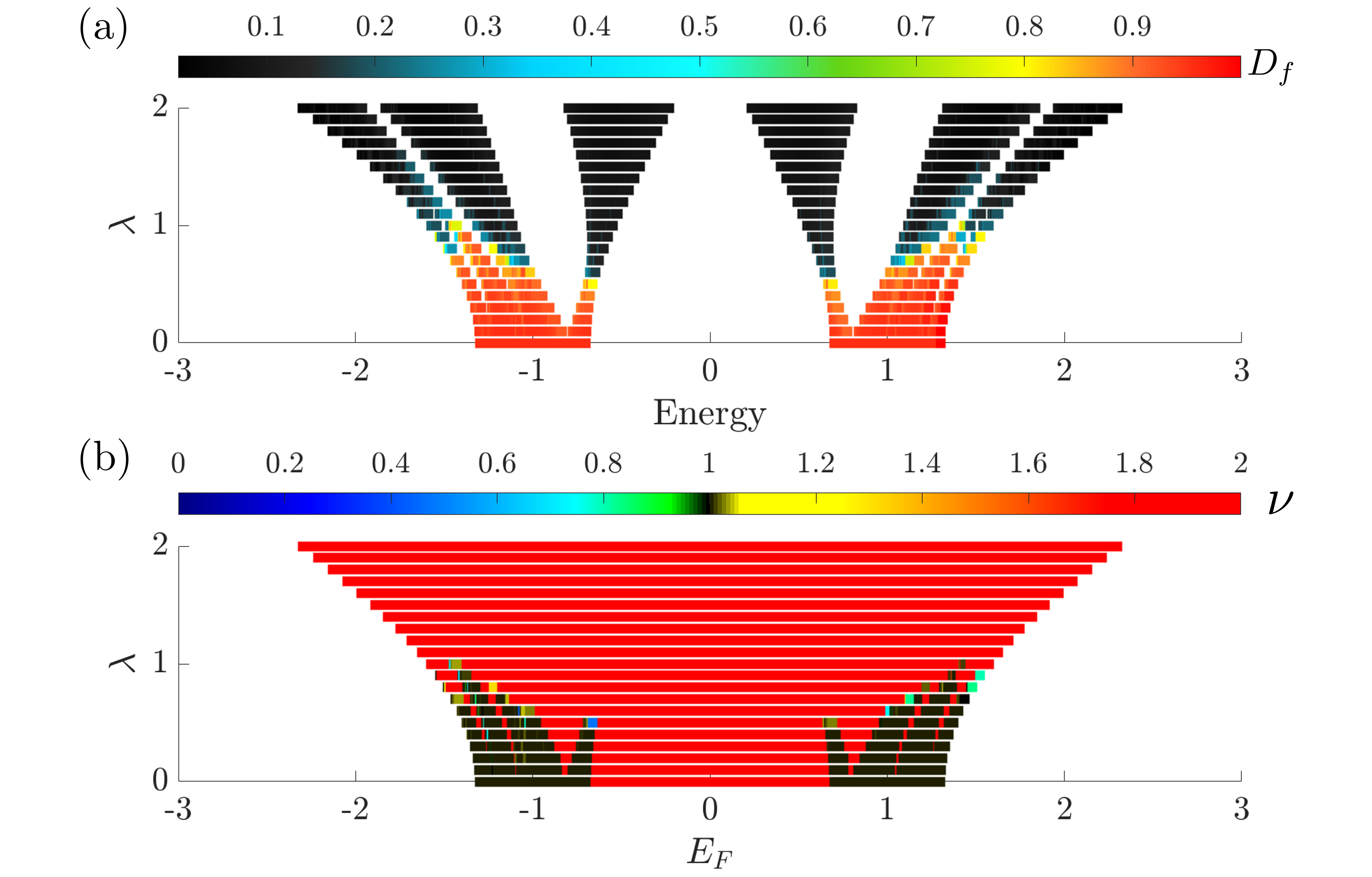}
    \caption{(a) Landscape of the finite-size fractal dimension of the eigenstates of the Hamiltonian in Eq.\eqref{AASSH}. (b) Scaling exponent of $S_Q$ for small-$q$, $\nu$ as the function of Fermi energy and $\lambda$. The system size is $N=10946$. $t_1=1$ and $t_2=0.3$.
    }
    \label{fig: suppleAASSH}
\end{figure}

\section{Detailed phase diagrams}
Figure~\ref{fig: supple} displays detailed phase diagrams of CHU, QHU and total hyperuniformity classes. Note that QHU (CHU) class I appears in the gap for extended (localized) regime.
QHU class III appears only in the gapless region at $\lambda=2$, and the class-III behavior remains in the total hyperuniformity, too.
\begin{figure}[h]
    \centering
    \includegraphics[width=1\textwidth]{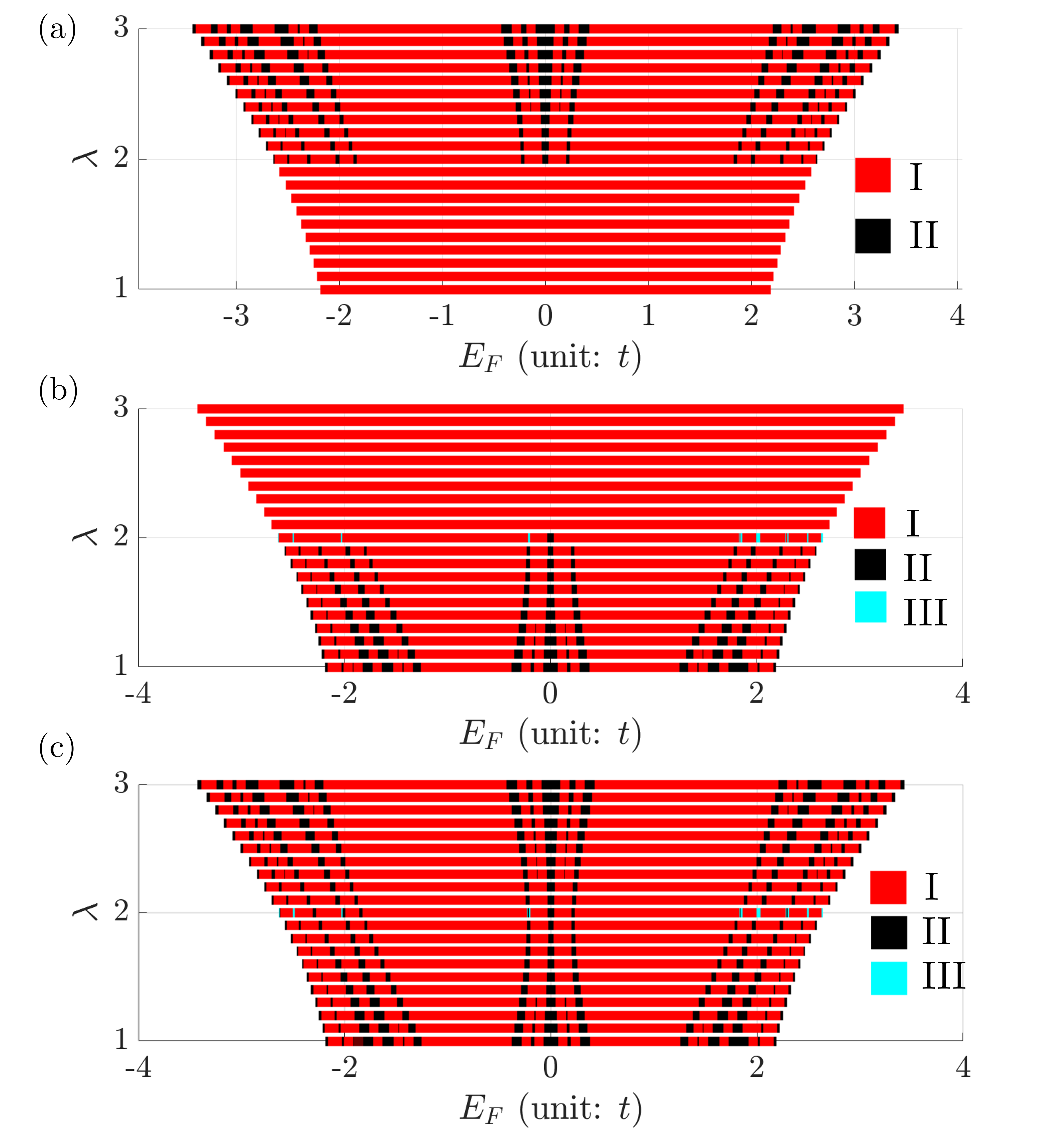}
    \caption{Phase diagrams of (a) CHU (b) QHU and (c) total hyperuniformity classes in the Aubry–André model. The system size is $N=17711$ and $t=1$.
    }
    \label{fig: supple}
\end{figure}

\section{Momentum-space wavefunction of the critical states}

In this section, we present the momentum-space wavefunction of the critical state. Note that the critical state is neither localized nor extended wavefunction. Due to the duality of the real and reciprocal spaces, the momentum-space wavefunction of the critical state is also neither localized nor extended. Figure~\ref{fig: momentumwavefunction} exhibits typical momentum-space wavefunction of critical state emergent in the Aubry–André model at $\lambda=2t$. Notably, critical states exhibit a fractal superposition of multiple momentum components, with enhanced amplitudes concentrated near specific wavevectors. Since adjacent states possess similar momentum components, the fractal structure of the momentum-space wavefunctions leads to an anomalous enhancement of small-$q$ quantum fluctuations through virtual particle-hole excitations based on Eq.~\eqref{mainresult}. This gives rise to the class-III QHU at some filling fractions.
\begin{figure}[h]
    \centering
    \includegraphics[width=0.8\textwidth]{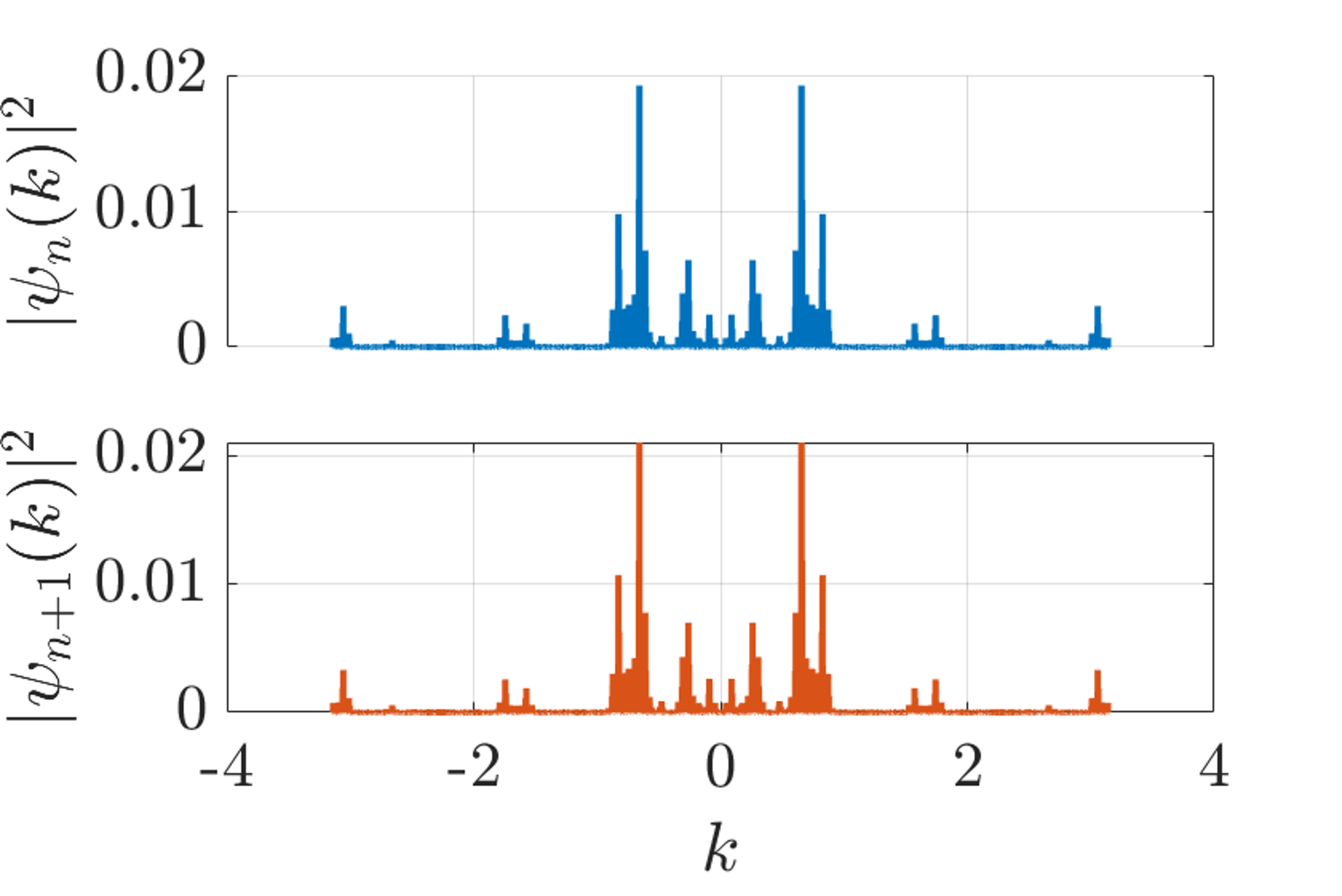}
    \caption{Probability distributions of momentum($k$)-space wavefunctions of adjacent critical states emergent in the Aubry–André model at $\lambda=2t$. Here, $n$ is eigenenergy index.
    }
    \label{fig: momentumwavefunction}
\end{figure}

\end{widetext}

\end{document}